# Solution-processed ZnO as the efficient passivation and electron selective layer of silicon solar cells

Jiangnan Ding[1], Yurong Zhou[1]*, Gangqiang Dong[1], Ming Liu[1,2], Donghong Yu[3,4], Fengzhen Liu[1,2]*

1. College of Materials Science and Opto-Electronic Technology, University of Chinese Academy of Sciences, 100049, Beijing, China
2. Sino-Danish college, University of Chinese Academy of Sciences, 100190, Beijing, China
3. Department of Chemistry and Bioscience, Aalborg University, DK-9220, Aalborg, Denmark
4. Sino-Danish Center for Education and Research (SDC), Niels Jensens Vej 2, DK-8000, Aarhus, Denmark

**Abstract:**

Solution-processed intrinsic ZnO and Al doped ZnO (ZnO:Al) were spin coated on textured n-type c-Si wafer to replace the phosphorus doped amorphous silicon as the electron selective transport layer (ESTL) of the Si heterojunction (SHJ) solar cells. Besides the function of electron selective transportation, the non-doped ZnO was found to possess certain passivation effect on c-Si wafer. The SHJ solar cells with different combinations of passivation layer (intrinsic a-Si:H, $SiO_x$ and non-doped ZnO) and electron transport layer (non-doped ZnO and ZnO:Al) were fabricated and compared. An efficiency up to 18.46% was achieved on a SHJ solar cell with an a-Si:H/ZnO:Al double layer back structure. And, the all solution-processed non-doped ZnO/ZnO:Al combination layer presents fairly good electron selective transportation property for SHJ solar cell, resulting in an efficiency of 17.13%. The carrier transport based on energy band diagrams of the rear side of the solar cells has been discussed related to the performance of the SHJ solar cells.

## 1 Introduction

Highly efficient solar cells can usually be achieved through a heterocontact or heterojunction concept, providing effective surface passivation and charge selective transportation. P-type a-Si:H/n-type c-Si heterojunction solar cells with interdigitated back contact (IBC) has achieved the record power conversion efficiency (PCE) of 26.3% for silicon solar cells with single absorber.[1] The main advantage of the a-Si:H/c-Si heterojunction structure lies in the excellent interface passivation brought by the intrinsic hydrogenated amorphous silicon (i a-Si:H).

In recent years, the silicon heterojunction concept has been expanded to wide band-gap materials which can generally be obtained using relatively simple techniques. High work function metal oxides (e.g. $MoO_x$, $WO_x$ and $V_2O_5$) were used as the hole selective transport layer (HSTL) on n type c-Si solar cells to substitute the p-type a-Si:H emitter.[2,3] The absorption losses brought by the doped a-Si:H can be much reduced owing to the low absorption coefficient of the metal oxides[3]. Recently, a PCE of 22.5% was achieved on $MoO_x$/n c-Si heterojunction solar cells[4]. At the same time, some wide band-gap materials with low work function were exploited as the electron selective transport layer (ESTL) to replace the n-type a-Si:H. For instance, using $TiO_x$ as the ESTL, efficiencies of 20.5%[5] and 21.6%[6] were achieved on n type c-Si heterojunction and homojunction solar cells respectively. Thermally evaporated $MgO_x$/Al was demonstrated to be an effective electron selective contact for homojunction silicon solar cells and enabling an efficiency of 20%[7]. Inspiring results have been achieved for novel Si heterojunction solar cells by making use of the wide band-gap materials as the HSL and ESTL at the same time. Efficiencies of 19.4% and 16.6% were respectively obtained on the dopant-free asymmetric heterocontacts c-Si solar cells with a $MoO_x$ layer as the HSTL and $LiF_x$ or a boron doped zinc oxide (BZO) as the ESTL[8,9].

To fully demonstrate the potential of such kind of solar cells, more heterocontact materials based on simple fabrication techniques need to be excavated and investigated. Low work function zinc oxide has the potential to serve as the ESTL of the SHJ solar cells and can be easily prepared by solution process. The work function and conductivity of ZnO can be easily adjusted by aluminum doping [10,11], which makes the application and optimization of ZnO ESTL more flexible. The solution process may ultimately lead to a significant reduction of the processing costs compared with various vacuum technologies such as atomic layer deposition (ALD) [5,6], metal-organic chemical vapor deposition(MOCVD)[9], etc. To our knowledge, however, the application of solution processed oxide as ESTL of n-type c-Si solar cells has not been explored.

In this paper, low work function zinc oxide thin films prepared by solution process were applied to the SHJ solar cells to serve as the ESTL. Novel SHJ solar cells with different combinations of interface passivation layer and ESTL were fabricated. An efficiency of 18.46% was achieved on a SHJ solar cell with an a-Si:H/ZnO:Al double layer back structure. And, an all solution-processed non-doped ZnO/ZnO:Al composition layer rear structure leads to an efficiency of 17.13% for the SHJ solar cell.

**2 Experimental details**

2.1 Synthesis of non-doped and Al-doped ZnO precursors

Non-doped ZnO precursor was synthetized using zinc acetate dihydrate ($ZnAc·2H_2O$, Analytical grade, China), ethanolamine (EA, Analytical grade, China), and dimethoxyethanol (DME, Analytical grade, China). One gram of $ZnAc·2H_2O$ was dissolved in 10 mL, 20 mL, and 40 mL of DME and made the Zn molar concentrations of 0.46M, 0.23M, 0.11M respectively. Then 280 μL EA was added into the solutions. After that, the mixtures were kept stirring for 12 hours at room temperature resulting various concentration of the ZnO precursors.

Synthesis of the Al-doped ZnO (ZnO:Al) precursors were similar to those of the non-doped ones, except that 10 mg aluminum nitrate ($Al(NO_3)_3 \cdot 9(H_2O)$) together with the one gram of $ZnAc \cdot 2H_2O$ was respectively dissolved in 10 mL, 20 mL, 40 mL and 60 mL of DME as the solvent for obtaining ZnO:Al precursors with different concentrations (Zn molar concentration: 0.46M, 0.23M, 0.11M and 0.08M).

2.2 Fabrication of the devices

N-type (100)-oriented CZ silicon wafers with a thickness of 220 μm and a conductivity of 1 to 5 Ω·cm were used as the substrates. After alkaline texturing, they were chemically cleaned according to the standard RCA procedure. Then, all of the substrates went through the same typical front side processes of SHJ solar cells. Intrinsic (~6 nm) and p-type (~10 nm) a-Si:H as the passivation and the emitter layer were successively deposited on the front side of the solar cells by means of plasma enhanced chemical vapor deposition (PECVD). A 80 nm thick indium tin oxide (ITO) layer was sputtered on top of the a-Si:H layer for electrical contact. Ag front grid was prepared by thermal evaporation.

To investigate the effect of the ESTL, 10 devices with different rear side structures were fabricated as depicted in Table 1. Non-doped ZnO and ZnO:Al precursors were respectively spin coated on the rear side of the devices 2#~4# and 5#~10# under 6000 rpm. The devices were then annealed at 200°C for 30 min for decomposition of the precursors and formation of non-doped ZnO or ZnO:Al as ESTL. Non-doped ZnO precursors used in devices 2#~4# have the same concentration (1g $ZnAc \cdot 2H_2O$ dissolved in 20mL DME). For devices 5#~8#, ZnO:Al precursors with different concentrations were used as shown in Table 1.

For devices 3#~8#, prior to the spin coating of ZnO or ZnO:Al precursors, the c-Si surface was passivated by using $SiO_x$ (3#) or intrinsic a-Si:H (4#-8#), which are two typical passivation thin films used in c-Si solar cells[12]. The $SiO_x$ layer was fabricated by $UV/O_3$ photo-oxidization. The a-Si:H thin film was deposited by using the PECVD technology under a substrate temperature of about 200°C. For device 9# with ZnO:Al ESTL, a non-doped ZnO (1g $ZnAc \cdot 2H_2O$ mixed in 40mL DME) layer is used as the passivation layer. The precursors of non-doped ZnO and ZnO:Al were spin coated and annealed successively to form the ZnO/ZnO:Al combination layer of device 9#. The sample 1# without ESTL but only a photo-oxidized $SiO_x$ layer at the rear side of the device is set as a reference.

Finally, a 400nm Al electrode was evaporated on the rear side of all of the devices. The active areas of the devices are about 1~1.5 cm$^2$.

2.3 Characterization

Surface morphology, microstructure and elemental analysis of the non-doped ZnO and ZnO:Al thin films on textured c-Si wafer were characterized by using scanning electron microscope (SEM, Hitachi SU8010), the backscattered electron microscope (BSE, Hitachi SU8010) and energy dispersive spectrometer (EDS, Bruker 6-30). Photocurrent density-voltage (J-V) curves of the solar cells were obtained under AM1.5 (100 mW/cm$^2$, 25°C) illumination. Thickness of the photo-oxidized $SiO_x$ layer on polished c-Si wafer was analyzed by spectroscopic ellipsometer (Sentech SENpro). Implied open circuit voltage (im$V_{oc}$) and pseudo fill factor (pFF) of the solar cells were obtained from Sinton Suns-$V_{oc}$ measurements.

Table 1 Photovoltaic parameters for reference cell (1#) and differently passivated but non-doped ZnO coated ones (2#-4#), a-Si:H passivated and doped ZnO coated devices with different precursor concentrations (5#-8#), and doped ZnO coated devices with and without ZnO as passivation layer (9#, 10#) . PL means passivation layer. ZAD, DME and AN represent $ZnAc·2H_2O$, dimethoxyethanol and $Al(NO_3)_3·9(H_2O)$, respectively.

| No. | PL | ESTL | ZAD:DME:AN (g);(mL);(mg) | $V_{oc}$/im$V_{oc}$[a] (mV) | $J_{sc}$ (mA/cm$^2$) | FF/pFF[b] (%) | Eff (%) | Rss (Ω) |
|---|---|---|---|---|---|---|---|---|
| 1# | SiO$_x$ | - | - | 544.4 | 34.83 | 73.78 | 13.99 | 3.51 |
| 2# | - | ZnO | 1:20:0 | 603.4 | 35.74 | 70.36 | 15.18 | 4.76 |
| 3# | SiO$_x$ | ZnO | 1:20:0 | 602.6 | 35.56 | 65.20 | 13.97 | 6.63 |
| 4# | a-Si:H | ZnO | 1:20:0 | 629.1 | 37.19 | 66.31 | 15.51 | 6.34 |
| 5# | a-Si:H | ZnO:Al | 1:10:10 | 642.1 | 34.63 | 60.82 | 13.52 | 5.84 |
| 6# | a-Si:H | ZnO:Al | 1:20:10 | 660.3/**661** | 37.68 | 65.09/**79.3** | 16.19 | 5.35 |
| 7# | a-Si:H | ZnO:Al | 1:40:10 | 655.0/**657** | 37.91 | 70.65/**78.8** | 17.54 | 4.50 |
| 8# | a-Si:H | ZnO:Al | 1:60:10 | 672.1/**675** | 38.23 | 71.95/**79.1** | 18.46 | 3.63 |
| 9# | ZnO | ZnO:Al | 1:40:10 | 645.3/**647** | 37.11 | 71.60/**78.9** | 17.13 | 3.97 |
| 10# | - | ZnO:Al | 1:40:10 | 559.4 | 36.08 | 73.44 | 14.81 | 3.35 |

[a] im$V_{oc}$ means implied $V_{oc}$. [b] pFF means pseudo FF.

## 3 Results and discussion

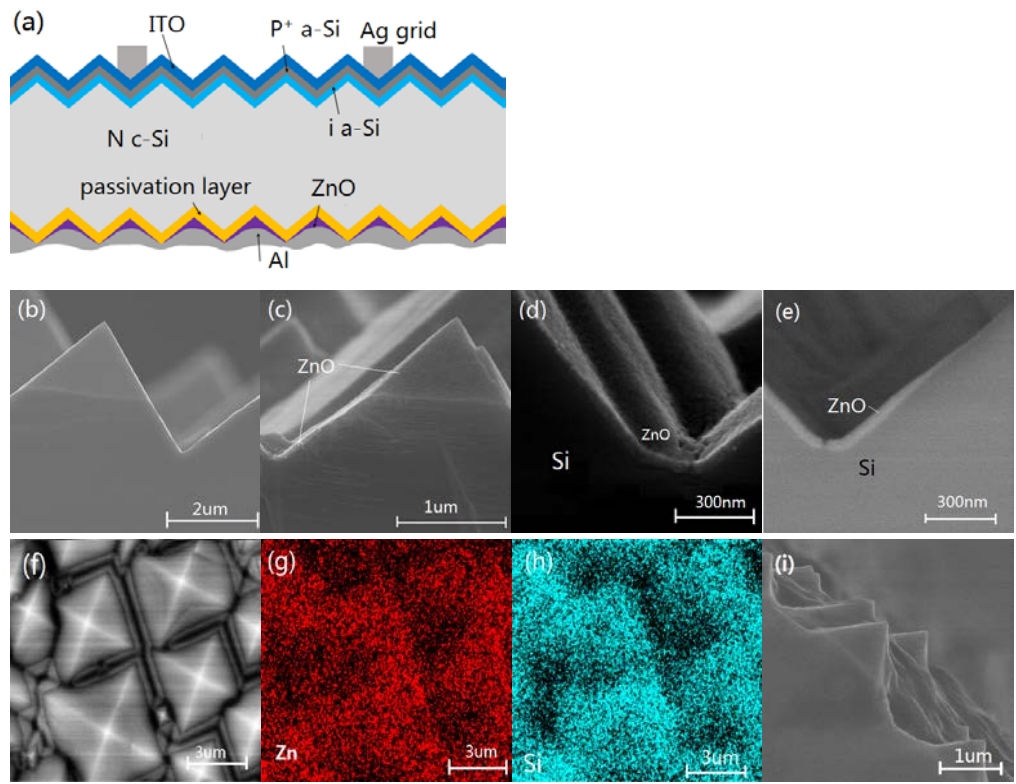

Figure1 (a) Cross-sectional schematic presentation of the SHJ solar cells with ZnO as ESTL; (b) Cross-sectional SEM image of a bare pyramidal-textured silicon wafer surface; (c) Cross-sectional SEM view of a region of ZnO coated pyramid-shaped Si surface ; (d) The close-up view of a valley of ZnO coated pyramid; (e) BSE image of a region of ZnO coated pyramid-shaped Si surface; (f) Top view of a zone of ZnO coated pyramid-shaped Si surface; (g) and (h) The corresponding EDS elemental mappings of Zn and Si, respectively; (i) SEM image of a region of heavily ZnO coated pyramid-shaped Si surface. The ZnO precursors were prepared with 1g ZnAc·2H$_2$O mixed in 20 mL DME for (c)-(h) and 1g ZnAc·2H$_2$O mixed in 10 mL DME for (i).

The configuration of the SHJ solar cells with ZnO as ESTL is illstrated in Figure 1a. In this paper, we mainly focus on the feasibility of the solution-processed ZnO as an electron selective layer of the SHJ solar cells. Therefore, the front side structure of all the devices studied in this paper is set as Figure 1a pronounced.

To be compatible with the commercial technology of Si solar cells, random pyramidal-texture was formed on both sides of the c-Si wafer via alkaline texturization, as what the typical cross-sectional SEM image presents in Figure 1b.

First of all, it is necessary to determine if the spin coated ZnO can form good coverage on the textured Si surface. The SEM image of a region of pyramid-shaped silicon surface covered with spin coated non-doped ZnO is shown in Figure 1c. The close-up of the pyramid bottom is depicted in Figure 1d. The amount of ZnAc·2H$_2$O and DME used to synthesize the ZnO precursor was 1 g and 20 mL, respectively. We can see from Figure 1c and 1d that the pyramid is covered completely with compact ZnO thin film, despite the non-uniform thickness of the ZnO layer at different positions of the pyramid. The thickness variation of the ZnO layer on the pyramidal-textured silicon wafer surface is more clearly presented in the BSE microscopy image as depicted in Figure 1e. On the upper part of the pyramid, the ZnO thickness is merely several nanometers. At the bottom of the pyramid, the thickness of the ZnO layer could be more than 50 nm. The SEM top view of a zone of pyramid-shaped Si covered with ZnO is shown in Figure 1f. The corresponding elemental EDS mappings of Zn and Si are given in Figure 1g, h, which confirm the full coverage of the pyramids by ZnO. Thickness inhomogeneity tends to happen when the films were formed on a rough substrate by spin coating of a precursor solution. And the thickness uniformity is also closely related to the precursor concentration. High concentration of the precursor is usually disadvantageous for the formation of uniform films. As shown in Figure 1i, the pyramids are half buried under the ZnO film when a sticky ZnO precursor (1g ZnAc·2H$_2$O mixed in 10 mL DME) was used. The thick ZnO layer at the bottom of the pyramids is detrimental to the carrier collection. Nevertheless, benefiting from a long diffusion length of the c-Si substrate, 30% surface coverage by ZnO with suitable thickness is enough to play the role of an effective ESTL of silicon solar cells[13]. According to the condition of Figure 1i, the surface coverage by thin ZnO layer (<10nm) can be estimated to be more than

50%. Therefore, using solution-processed ZnO as the ESTL of Si solar cells is feasible.

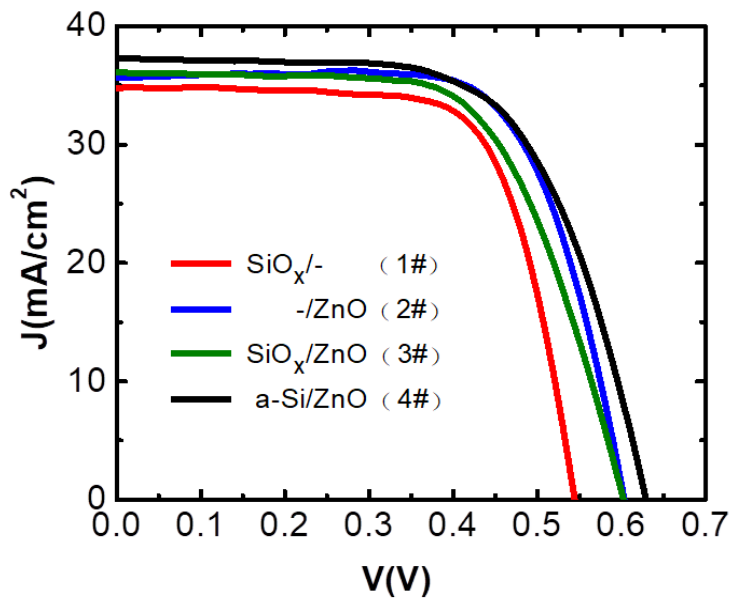

Figure 2 *J–V* characteristics of the SHJ solar cells with non-doped ZnO ESTL. The passivation layers are $SiO_x$ and a-Si:H for sample 3# and 4# respectively. No passivation layer was used for device 2#. The reference solar cell (1#) without ZnO ESTL but only a photo-oxidized $SiO_x$ layer at the rear side of the device is also presented.

The J-V curves of the SHJ solar cells with spin coated non-doped ZnO as the ESTL and with different passivation conditions are revealed in Figure 2. For comparison, J-V curve of the reference device (1#), without ZnO back contact layer but only a photo-oxidized $SiO_x$ passivation layer at the rear side of the solar cell, is included in Figure 2.

For the reference device (1#), nice Ohmic contact is formed between the Al back electrode and the n-type c-Si substrate with the help of a thin $SiO_x$ passivation layer (8 min photo-oxidation, 1.5 nm $SiO_x$ layer on polished Si). The analysis about the Ohmic contact is provided in supporting information (Supplementary 1). An open circuit voltage ($V_{oc}$) of about 544 mV and an efficiency of about 14% were obtained for the reference device with such a simple back structure. Similar Ohmic contact was observed when $MgO_x$ was used as the passivation layer of c-Si[7], where $V_{oc}$ of 628mV and PCE of 20% were obtained.

If we compare the devices 2# and 3# with the reference device 1#, we can find that, whether or not there is a $SiO_x$ passivation layer, the $V_{oc}$ of the solar cells with a ZnO ESTL layer (2# and 3#) are all about 60 mV higher than that of the reference device. This implies that the ZnO ESTL does play an important role in carrier selective transportation. However, the $V_{oc}$ of device 2# and 3# are still significantly lower than that of a traditional SHJ solar cell with a-Si:H ESTL[1]. Considering the excellent passivition effect of the intrinsic a-Si:H and its critical role in increasing $V_{oc}$, a 6 nm intrinsic a-Si:H layer was used as the passivation layer of the silicon wafer before the ZnO layer fabrication (device 4#). And the $V_{oc}$ of device 4# is further improved to 629mV as shown in Table 1 and Figure 2.

The photovoltaic parameters of the SHJ solar cells with non-doped or doped ZnO as ESTL and differently passivated or non-passivated layers are summarized in Table 1. We noticed that inserting a $SiO_x$ passivation layer between the c-Si wafer and the non-doped ZnO ESTL does not improve the $V_{oc}$ of the solar cell (compare 2# with 3#). The phenomenon proves that the ZnO layer itself exhibits a certain passivation ability on the c-Si surface. And the passivation effect of ZnO is identical to that of the photo-oxidized $SiO_x$. From all improvements on $V_{oc}$, short circuit current density ($J_{sc}$), and fill factor (FF), we can find that, as a back contact of the SHJ solar cells, the combination layer of a-Si:H/ZnO (4#) functions much better than the $SiO_x$/ZnO double layer (3#). Such results conclude the excellent passivation effect of the intrinsic a-Si:H, which is better than that of the photo-oxidized $SiO_x$.

Compare the parameters of devices 2#, 3#, and 4# with that of device 1#, it is easy to find that the addition of the ZnO layer increases the series resistance and then lowers the FF of the devices as shown in Table 1. This should be attributed to the relatively low conductivity and large thickness of the non-doped ZnO layer. Therefore, aluminium doped ZnO precursors with different concentrations were further suggested to prepare the ESTL.

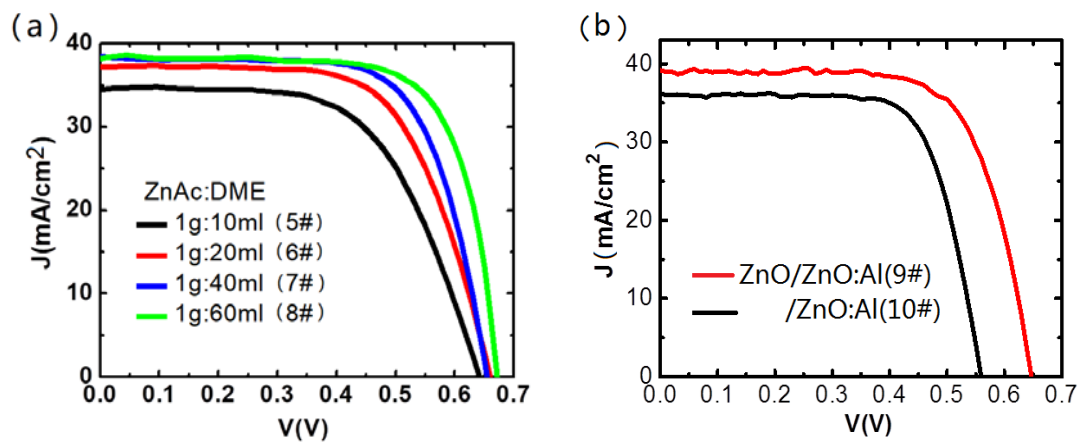

Fig. 3 *J–V* characteristics of the different SHJ devices with ZnO:Al as ESTL: (a) ZnO:Al precursor under different concentrations for ESTL formation with a-Si:H passivation layer; (b) SHJ solar cells with all solution-processed ZnO/ZnO:Al combination layer back contact (9#) and with only ZnO:Al ESTL (10#).

Figure 3a shows the *J–V* characteristics of the devices 5#~8# prepared with different ZnO:Al precursor concentrations. Aluminum doping improves the conductivity of ZnO:Al ($1.3×10^{-5}$ S/cm) by about 2 orders of magnitude compared with that of the non-doped ZnO ($2.5×10^{-7}$ S/cm), which leads to lower series resistance of this series of devices, as listed in Table 1. Compared with the device 4# (a-Si:H/ZnO), ZnO:Al as ESTL increases the $V_{oc}$ of the devices 5#~8# (a-Si:H/ZnO:Al) evidently. In addition, the decrease of the precursor concentration of ZnO:Al improves the carrier collection and the device performance gradually. The highest efficiency of 18.46% (device 8#) with $V_{oc}$ of 672mV, $J_{sc}$ of 38.2 mA/cm$^2$ and FF of 71.9% was achieved for a SHJ solar cell with an a-Si:H/ZnO:Al combination layer under the lowest precursor concentration. The electron transport process will be discussed later with the assistant of the energy band structures.

In order to further simplify the fabrication process and taking in to account the fairly nice passivation effect of ZnO at the same time, solution-processed non-doped ZnO was also used as a passivation layer for the ZnO:Al ESTL (device 9#). For comparison, a device without the non-doped ZnO passivation layer but only the ZnO:Al ESTL was prepared (device 10#).

Figure 3b shows the *J–V* characteristics of devices 9# and 10#. The $V_{oc}$ of device 10# is barely improved compared with that of the reference device 1#. It could be due to weakening of the passivation effect of ZnO by Al doping. It is encouraging that the performance of device 9# with the all solution-processed non-doped ZnO/ZnO:Al back structure is very close to those of the SHJ solar cells with a-Si:H/ZnO:Al combination layers. An efficiency of 17.13% (device 9#) with $V_{oc}$ of 645 mV, $J_{sc}$ of 37.1 mA/cm$^2$ and FF of 71.6% was achieved for the SHJ solar cell with a ZnO/ZnO:Al back contact. To the best of the authors' knowledge, there is no such ESTL structure reported. The absorption coefficients of ZnO and ZnO:Al are much lower than that of a-Si:H[14]. They are more appropriate than a-Si:H to serve as the ESTL in a Si solar cell with bifacial structure which has been widely adopted recently[15].

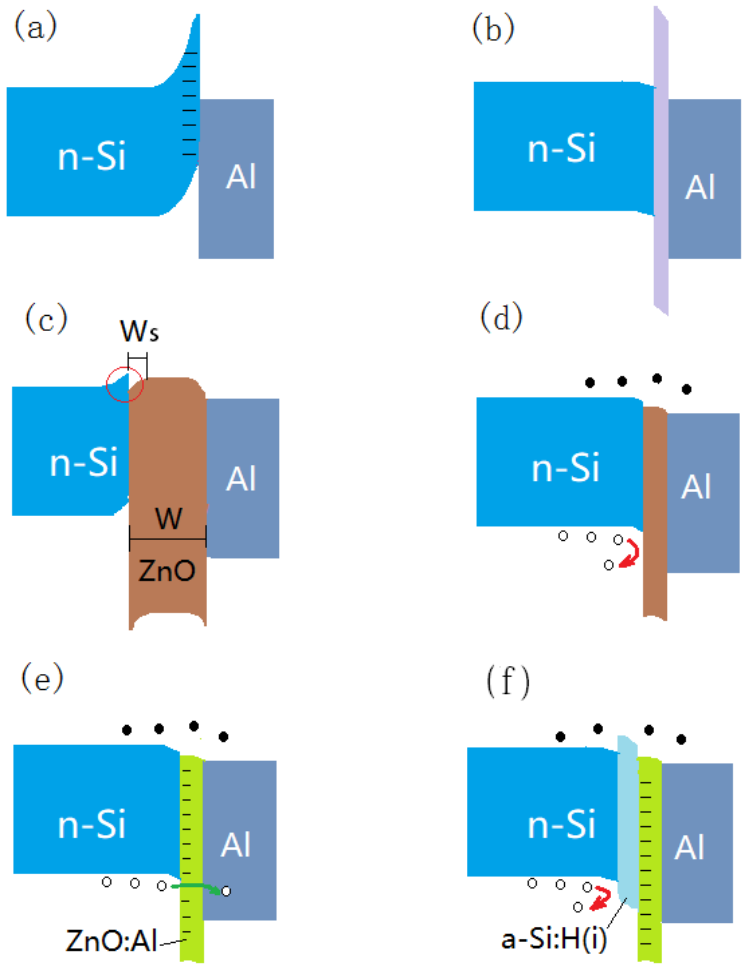

Fig.4 Schematic energy band diagrams of the rear side of the SHJ devices with different structures. (a) n-Si/Al, (b) n-Si/SiO$_x$/Al, (c) n-Si/ZnO (thick)/Al, (d) n-Si/ZnO (thin)/Al, (e) n-Si/ZnO:Al/Al, (f) n-Si/a-Si:H/ZnO:Al/Al.

To understand the behavior of the above SHJ solar cells, schematic energy band diagrams of the rear contact with different structures are displayed in Figure 4. The work functions of Al, ZnO, and n-Si are set to be 4.2 eV, 4.45 eV[16], and 4.25 eV, respectively. The electron affinities of Si and ZnO are set to be 4.05eV and 4.2eV[17], respectively. Without effective surface passivation, a Schottky-barrier with the height of about 0.65 eV[18] will be formed at the n-Si/Al interface due to the existence of high density surface states of the Si wafer, as shown in Figure 4a. As the Si surface is passivated by photo-oxidized SiO$_x$ (device 1#), the high Schottky-barrier vanishes and an ohmic contact is formed (Figure 4b). In this case, both electrons and holes can go through the ultrathin SiO$_x$ layer by tunneling. However, the electrons and holes cannot be effectively separated at the back contact. As a result, recombination of the photogenerated carriers at the rear side of the device leads to poor performance.

If a thick ZnO layer is used to replace the SiO$_x$ layer, the ideal energy band diagram should be like Figure 4c. The offset between ZnO and Si on conduction band (CB) is about 0.15 eV. Although the CB offset is not very large, such structure is disadvantageous for carriers' transportation. If the thickness of the ZnO layer (W) is far less than that of the space-charge region ($W_s$), charge exchange will happen between the Al contact and the c-Si substrate. The up-bending part of the energy band of the Si surface in the circle of Figure 4c will bend down, as demonstrated in Figure 4d. In this case, electrons can go through the contact smoothly. However, holes will be blocked by the large barrier in the valence band (VB). So, the thin ZnO layer functions as an effective electron selective layer and the $V_{oc}$ of device 2# is evidently improved in comparison with the reference device 1#.

When ZnO:Al is applied as the ESTL, tunnel of holes through the defects in the VB barrier (Figure 4e) probably leads to low $V_{oc}$ of device 10#. When intrinsic a-Si:H or ZnO is inserted between the c-Si substrate and the ZnO:Al layer (Figure 4f), tunneling of holes is blocked. As a result, the $V_{oc}$ of devices 5#~9# are improved significantly.

According to the schematic energy band diagrams, small thickness and large valence band offset of the ESTL play an important role in the effective carrier collection of n-type c-Si solar cell. With an Al electrode, the work function of the ESTL material does not have to be lower than that of the n-type Si.

It is necessary to point out that the fabrications of the front side ITO and all the metal contacts of the devices were not well optimized. Both ZnO and ZnO:Al are amorphous according to the XRD measurements (Supplementary 3). As a result, the series resistances of all of the solar cells are quite high, which limit the efficiency of the devices. To eliminate the influence of the series resistance, Suns-Voc measurements were carried out on devices 6#~9#. The implied $V_{oc}$ and pseudo FF are provided in Table 1. We can see that the implied $V_{oc}$ are consistent with the actual measurements. However, the pseudo FFs are much higher than the actual values. Considering a $J_{sc}$ value of 39 mA/cm$^2$ for a conventional Si solar cell, the pseudo efficiency for devices 6#~9# could be above 20%. So, we can predict that through further optimization of the ITO and the front and rear metal contacts, an efficiency of above 20% is very hopeful for the SHJ solar cells with low-cost solution processed ZnO ESTL.

4 **Conclusions**

Solution-processed ZnO or ZnO:Al were used as the electron selective contact for silicon heterojunction solar cells. Different concentration of ZnO or ZnO:Al precursors and different interface passivation materials were applied. Efficiency up to 18.46% with a $V_{oc}$ of 672 mV was achieved for a SHJ solar cell with an a-Si:H/ZnO:Al combination electron selective layer. It was also demonstrated that ZnO has a certain passivation effect on silicon wafers. As a result, 17.13% efficiency was achieved on a SHJ solar cell with all solution-processed ZnO/ZnO:Al combination ESTL layer.

The low-cost solution-processed ZnO electron selective materials could meet both of the optical and electrical requirements of a SHJ solar cell. Meanwhile the process temperatures for the whole devices are not higher than 200 °C. So, this work not only highlights the potential of the solution-processed ZnO in high-efficiency and low-cost silicon solar cells, but also provides an alternative concept for designing and fabricating other optoelectronic devices.

**Acknowledgments**

This work was supported by the National Natural Science Foundation of China (No.61604153 and No.61674150). Support from the Sino-Danish Center for Education and Research (SDC) is also acknowledged.

Reference
1. Yoshikawa K, Kawasaki H, et al, Silicon heterojunction solar cell with interdigitated back contacts for a photoconversion efficiency over 26%. *Nature Energy* 2017;2:17032
2. M. Bivour, J. Temmler, H. Steinkemper and M. Hermle, *Solar Energy Materials and Solar Cells*, 2015, **142**, 34-41.
3. C. Battaglia, S. M. De Nicolas, S. De Wolf, X. Yin, M. Zheng, C. Ballif and A. Javey, *Applied Physics Letters*, 2014, **104**, 113902.
4. J. Geissbühler, J. Werner, S. Martin de Nicolas, L. Barraud, A. Hessler-Wyser, M. Despeisse, S. Nicolay, A. Tomasi, B. Niesen and S. De Wolf, *Applied Physics Letters*, 2015, **107**, 081601.
5. X. Yang, P. Zheng, Q. Bi and K. Weber, *Solar Energy Materials and Solar Cells*, 2016, **150**, 32-38.
6. X. Yang, Q. Bi, H. Ali, K. Davis, W. V. Schoenfeld and K. Weber, *Advanced Materials*, 2016, **28**, 5891-5897.
7. Y. Wan, C. Samundsett, J. Bullock, M. Hettick, T. Allen, D. Yan, J. Peng, Y. Wu, J. Cui, A. Javey, A. Cuecas, *Advanced Energy Matererials,* 2017, **7**, 1601863.
8. J. Bullock, M. Hettick, J. Geissbühler, A. J. Ong, T. Allen, C. M. Sutter-Fella, T. Chen, H. Ota, E. W. Schaler, S. De Wolf, C. Ballif, A. Cuevas, A. Javey, *Nature Energy,* 2016, **1**, 15031.
9. F. Wang, S. Zhao, B. Liu, Y. Li, Q. Ren, R. Du, N. Wang, C. Wei, X. Chen and G. Wang, *Nano Energy*, 2017, **39**, 437-443.
10. T. Tsuchiya, T. Emoto and T. Sei, *Journal of Non-Crystalline Solids*, 1994, **178**, 327-332.
11. W. Tang and D. Cameron, *Thin solid films*, 1994, **238**, 83-87.
12. F. Feldmann, M. Bivour, C. Reichel, M. Hermle and S. W. Glunz, *Solar Energy Materials and Solar Cells*, 2014, **120**, 270-274.
13. E. Franklin, K. Fong, K. McIntosh, A. Fell, A. Blakers, T. Kho, D. Walter, D. Wang, N. Zin and M. Stocks, *Progress in Photovoltaics: research and applications*, 2016, **24**, 411-427.
14. D. Song, A. G. Aberle and J. Xia, *Applied Surface Science*, 2002, **195**, 291-296.

15. T. Mishima, M. Taguchi, H. Sakata and E. Maruyama, *Solar Energy Materials and Solar Cells*, 2011, **95**, 18-21.
16. G. Murdoch, S. Hinds, E. Sargent, S. Tsang, L. Mordoukhovski and Z. Lu, *Applied Physics Letters*, 2009, **94**, 138.
17. S. M. Sze and K. K. Ng, *Physics of semiconductor devices*, John wiley & sons, 2006.
18. D. Baik and S. Cho, *Thin Solid Films*, 1999, **354**, 227-231.